\documentclass{PoS}

\title{Chiral magnetic effect in 2+1 flavor QCD+QED}

\ShortTitle{Chiral magnetic effect in 2+1 flavor QCD+QED}

\author{M. Abramczyk\\
        E-mail: \email{mabramc@gmail.com}}
\author{\speaker{T. Blum}\\
        E-mail: \email{tblum@phys.uconn.edu}}
\author{G. Petropoulos\\
        E-mail: \email{gregpetrop@gmail.com}}
\author{R. Zhou, \thanks{We thank RIKEN and the US DOE for providing the resources to complete this work.}\\
E-mail: \email{zhouran123@gmail.com}\\	
        Physics Department, University of Connecticut, 2152 Hillside Rd., Storrs, CT, 06269-3046, USA
        }

\abstract{The exciting possibility of direct observation of QCD instantons 
in heavy-ion collisions has recently been proposed by Kharzeev.  
The underlying phenomenon, known as the chiral magnetic effect,
may have been observed recently at RHIC, and a first principles calculation 
is needed to confirm and understand the results. 
The chiral magnetic effect is thought to be visible in the symmetric phase, 
at temperatures above the QCD critical temperature, and in the 
presence of an external magnetic field. We report on first 
2+1 flavor, domain wall fermion, QCD+QED dynamical simulations 
above the critical temperature, in a fixed topological sector(s),
which are used to study the electric charge separation produced by the effect.}

\FullConference{The XXVII International Symposium on Lattice Field Theory - LAT2009\\
		 July 26-31 2009\\
		 Peking University, Beijing, China}

\newcommand{\dslash}{\rlap{/}\kern-2.0pt \partial}
\newcommand{\Dslash}{\rlap{/}\kern-2.5pt D}

\begin{document}

\section{Introduction}

Kharzeev proposed that topological charge fluctuations can be observed in the quark-gluon plasma created in heavy-ion collisions in the presence of a strong magnetic field~\cite{Kharzeev:2004ey}. The topological charge fluctuations induce an effective $\theta$ parameter with a spatial gradient which couples to the magnetic field to separate positive and negative electric charge\cite{Kharzeev:2007tn}. The effect, dubbed the chiral magnetic effect, has reportedly been observed by the STAR experiment at RHIC\cite{Selyuzhenkov:2005xa,Voloshin:2008jx}. 
Because the effect depends on the strength of the magnetic field, and its observation on the system being in the chirally symmetric phase~\cite{Kharzeev:2007jp,Fukushima:2008xe}, it is important to study the chiral magnetic effect in the framework of lattice QCD in order to confirm this exciting discovery and understand it in detail. The Moscow group reported at the meeting on a study in quenched $SU(2)$ gauge theory\cite{Buividovich:2009wi}. 

The physical picture used and referred to throughout this talk is the following. Begin with a region where the gluon field strength times its dual, $G\tilde G$, is non-zero. In the classical picture, this region corresponds to an instanton, and there will be a chiral zero mode associated with it. This zero-mode is an equal mixture of quark and anti-quark components. When the magnetic field $\vec B$ is applied over this region, the zero-mode is polarized along the direction of the magnetic field, with the quark residing slightly more on one hemisphere and the anti-quark on the other. Thus, there is a small polarization of electric charge near the ``surface" of the instanton.  Because the relevant quark degrees of freedom are the low-lying modes of the Dirac operator, it is useful to work with its spectral decomposition, and for practical reasons we work with the Hermitian Dirac operator  $\Dslash_H= \gamma_5\,\,\Dslash_H$. 
The charge density in units of $e$ in the continuum then becomes 
\begin{eqnarray*}
\rho &=& \bar\psi\gamma_0\psi=~  i{\rm tr}\,\gamma_5\,\,\Dslash_H\gamma_4
=i \sum_\lambda \frac{\psi_\lambda^\dagger\gamma_4\gamma_5\psi_\lambda}{\lambda+m}
\end{eqnarray*}
where the eigenvalues $\lambda$ are real.  The extra factor of $i$ is necessary to continue back to Minkowski space. A similar expression with similar properties holds in the case of domain wall fermions (DWF) at non-zero lattice spacing $a$\cite{Blum:2001qg}. 
It is easy to see that $\rho=0$ for an exactly chiral mode, so that after polarization by the external field, the would-be zero-mode has neither zero eigenvalue nor is it exactly chiral. However, it remains nearly chiral as we shall see.
Strictly speaking, the lowest modes of the DWF Dirac operator are not exact zero-modes, but near zero-modes because of the small, explicit, chiral symmetry breaking induced by finite $L_s$. Here we treat this small breaking as negligible compared to that induced by the magnetic field, but this still needs to be demonstrated.

\section{Simulation details}
\begin{figure}[hbt]
\centerline{
  \includegraphics[scale=0.3]{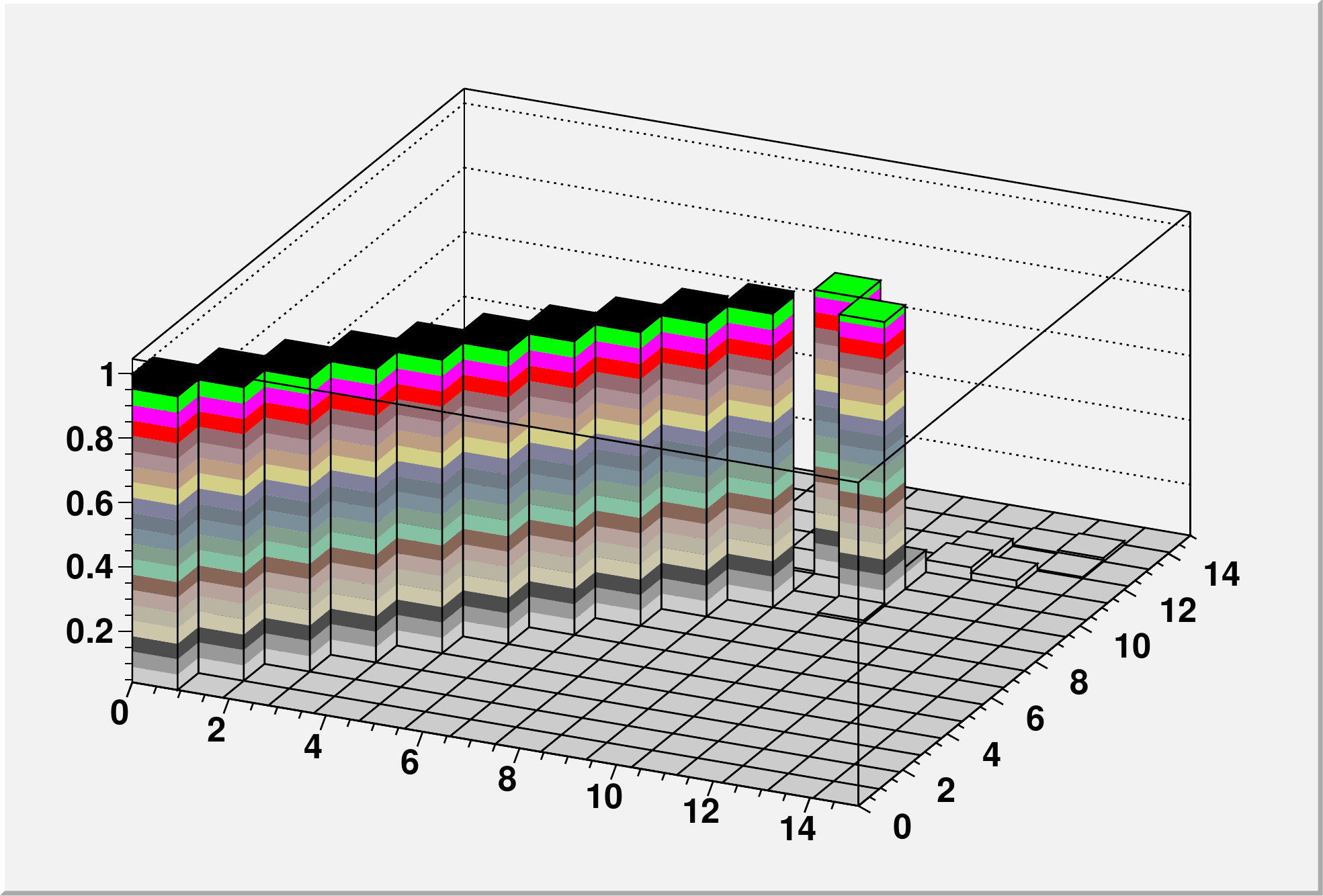} 
 } 
  \caption{Chirality $\langle\psi_i|\Gamma_5|\psi_j\rangle$ of the low-modes of the DWF Dirac operator for the QCD configuration 420 discussed in the text, with $B_z=0$. There are 10 zero-modes in this configuration, so the topological charge is $Q=10$ by the Atiyah-Singer index theorem.}
\label{fig:chirality}
\end{figure}

To study the chiral magnetic effect in 2+1 flavor QCD at non-zero temperature and topological charge, we use DWF on a lattice of size $16^3\times 8$, the RHMC algorithm of Clark and Kennedy, and an auxiliary determinant (to fix the topological charge)\cite{Renfrew:2009wu,Vranas:2006zk,Fukaya:2006vs}. The Iwasaki gauge coupling is $\beta_{QCD}=1.80$, extra dimension size $L_s=16$, light quark mass $m_l=0.013$, and strange quark mass $m_s=0.04$. The twisted masses in the auxiliary determinant are $\epsilon_f=0.0001$ and $\epsilon_b=0.50$ (see Ref.~\cite{Renfrew:2009wu} for definitions). Our simulation parameters were chosen so the system is above the critical temperature, $T_c$.
To fix the topological charge, the evolution was started with $\epsilon_f=0.5$ and gradually reduced to its final value to ``freeze" the topological charge to a value of $Q=9$ or 10. Though it is a secondary effect, to have the complete electromagnetic picture, we also couple the dynamical quarks to photons as well as the external magnetic field. The QED Wilson gauge action is used, with coupling $\beta_{QED}=1.5$. For simplicity for now, the quark charges are set equal. The flux of the external field is quantized because of the periodic boundary conditions in units of $2 \pi/ q L^2$ \cite{AlHashimi:2008hr,Detmold:2009dx} ($q$ is quark charge), and to maintain gauge invariance, the quarks obey a twisted boundary condition~\cite{AlHashimi:2008hr}.
For measurements of the eigenvectors of the Dirac operator, a valence quark value of $L_s=32$ was used in order to obtain low-modes with continuum-like chiral properties (see Fig.~\ref{fig:chirality}, and for  a detailed discussion of the DWF eigenvalue spectrum and how it is computed, Ref.~\cite{Blum:2001qg}). 
%\begin{minipage}{25pc}
%\end{minipage}

\section{Results}

\subsection{Continuum-like instanton}

\begin{figure}[hbt]
  \includegraphics[scale=0.2]{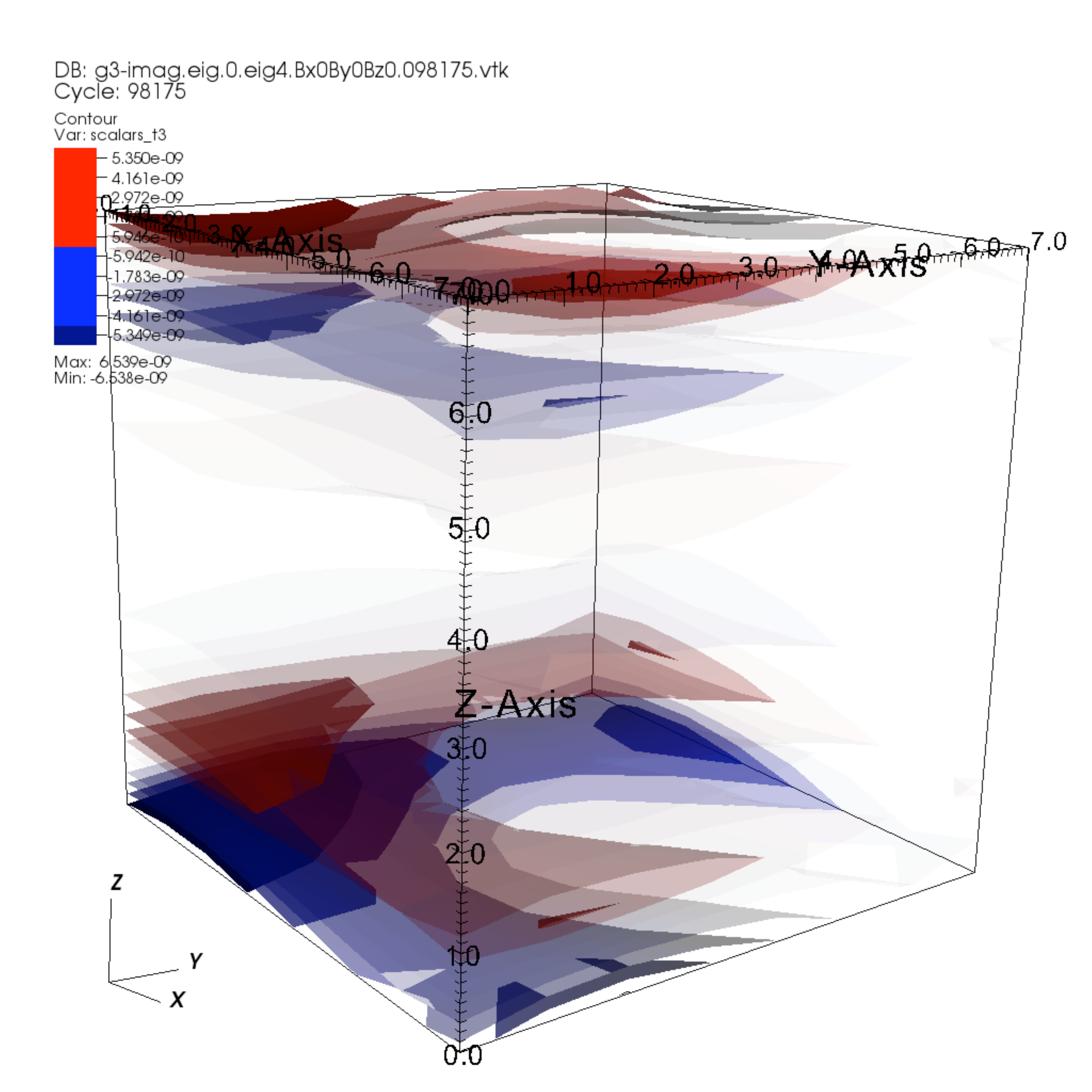}
  \includegraphics[scale=0.35]{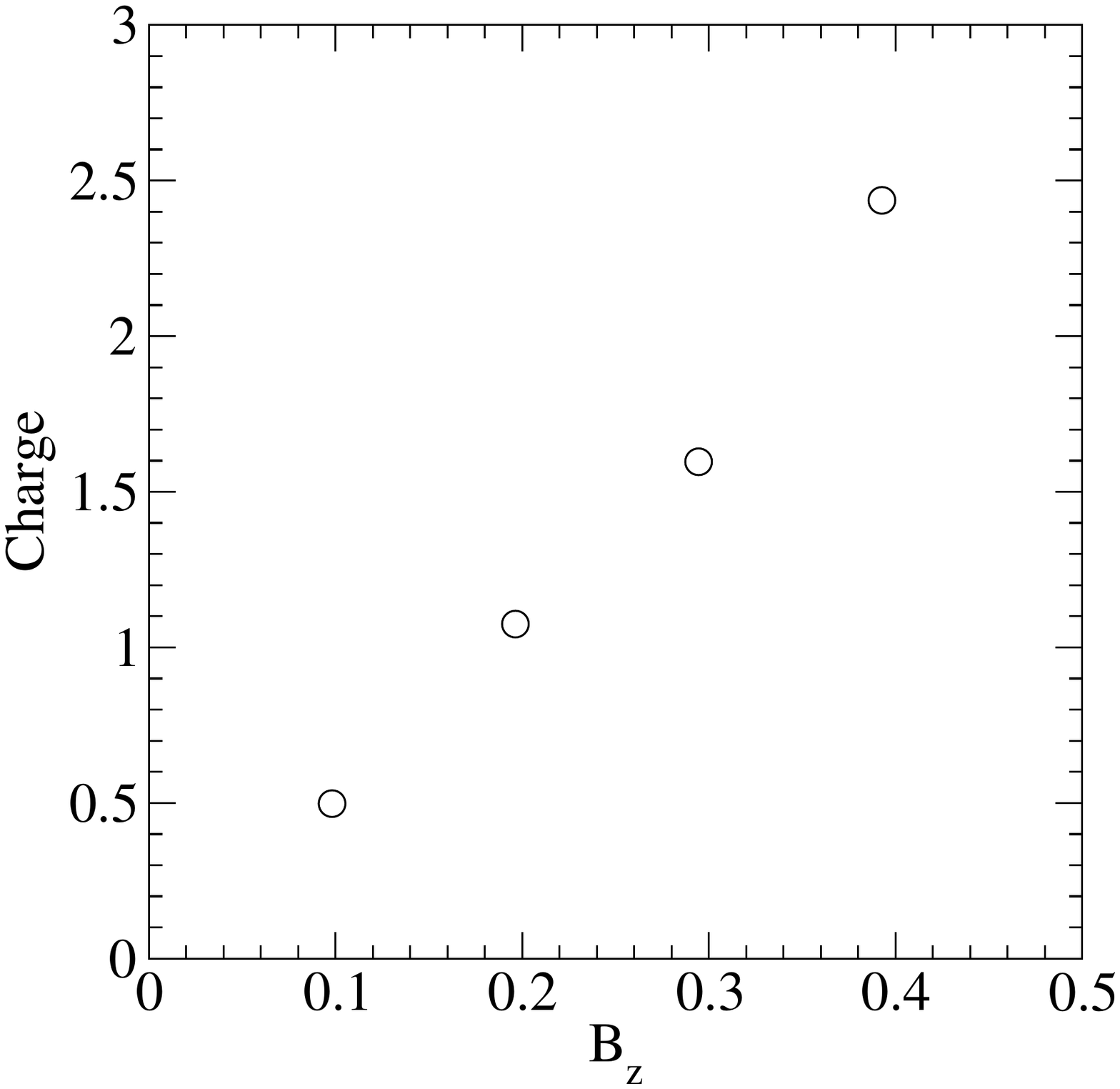}
  \caption{Left panel: Charge separation computed from a single near-zero-mode for a continuum instanton discretized on an $8^4$ lattice. $B_z=0.098175$. Translational invariance is broken in the $x-y$ plane by the Landau states of the quarks. Right panel: total amount of charge separated to the lower half of the lattice in the $z$ direction for the same configuration. 
All modes with chirality close to one are included in the total.
The same amount, but with opposite sign resides in the top half.}
\label{fig:charge sep 1}
\end{figure}

Before proceeding to the phenomenologically  interesting case of 2+1 flavor QCD above $T_c$,
we investigate the chiral magnetic effect in the background of a single instanton that has been discretized onto a lattice\cite{Chen:1998ne}  with size $8^4$ sites.
\begin{eqnarray}
A_\mu&=&-i\sum_{j=1}^3\eta^{j\mu\nu}\lambda_j\frac{x_\nu}{x^2+\rho^2}\\
\rho(r) &=& \rho_0\left(1-\frac{r}{r_{\rm max}}\right)\Theta(r_{\rm max}-r)
\end{eqnarray}
where the continuum instanton of size $\rho_0/a=10$ is smoothly cut off at a distance from the center of $r_{max}/a=3$. Because it is cut off, the instanton supports several zero-modes, not just one. In the presence of a non-zero magnetic field pointing along the $z$-direction, layers of positive and negative charge appear above and below  the instanton in this direction. In Fig.~\ref{fig:charge sep 1}, this separation is shown for the first zero-mode associated with the instanton. Note, both positive(red) and negative(blue) charge appears above and below. However, there is a $net$ positive charge above and minus that below so that the total charge on the configuration is zero, as it must be. Whether the dipole structure above and below the instanton is a physical effect~\cite{Buividovich:2009my}, an artifact of the discretization of the continuum instanton, or both, is not clear and needs further investigation.
Next, the magnetic flux is doubled, tripled, and quadrupled, and the total charge computed above and below the mid-plane of the lattice in the $z$-direction. The results are plotted in the right panel of Fig.~\ref{fig:charge sep 1}. An almost linear rise with the strength of the magnetic field is found when only those eigenvectors that are nearly chiral, $\langle \lambda_i | \Gamma_5|\lambda_j\rangle\approx1$ are used in the charge sums. Higher modes which have chirality significantly less than one, but are not paired, are presumably lattice artifacts; paired modes do not contribute.
The breaking of translation invariance in the $x-y$ plane due to the quark Landau levels, which would happen as well in the absence of the instanton\cite{AlHashimi:2008hr}, is clearly visible in Fig.~\ref{fig:charge sep 1}.

\subsection{2+1 flavor QCD}

\begin{figure}[hbt]
  \includegraphics[scale=0.22]{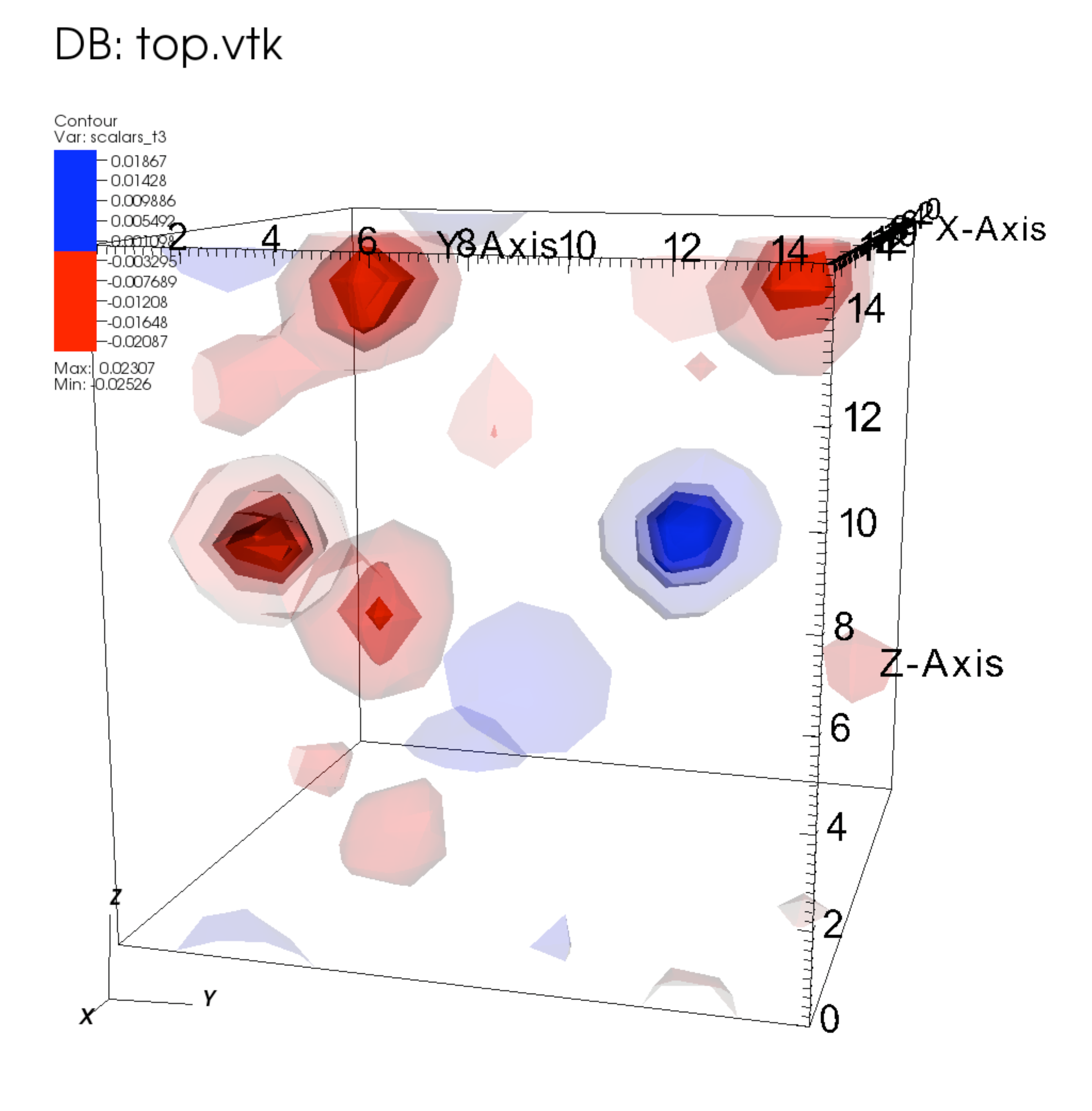}
  \includegraphics[width=18pc]{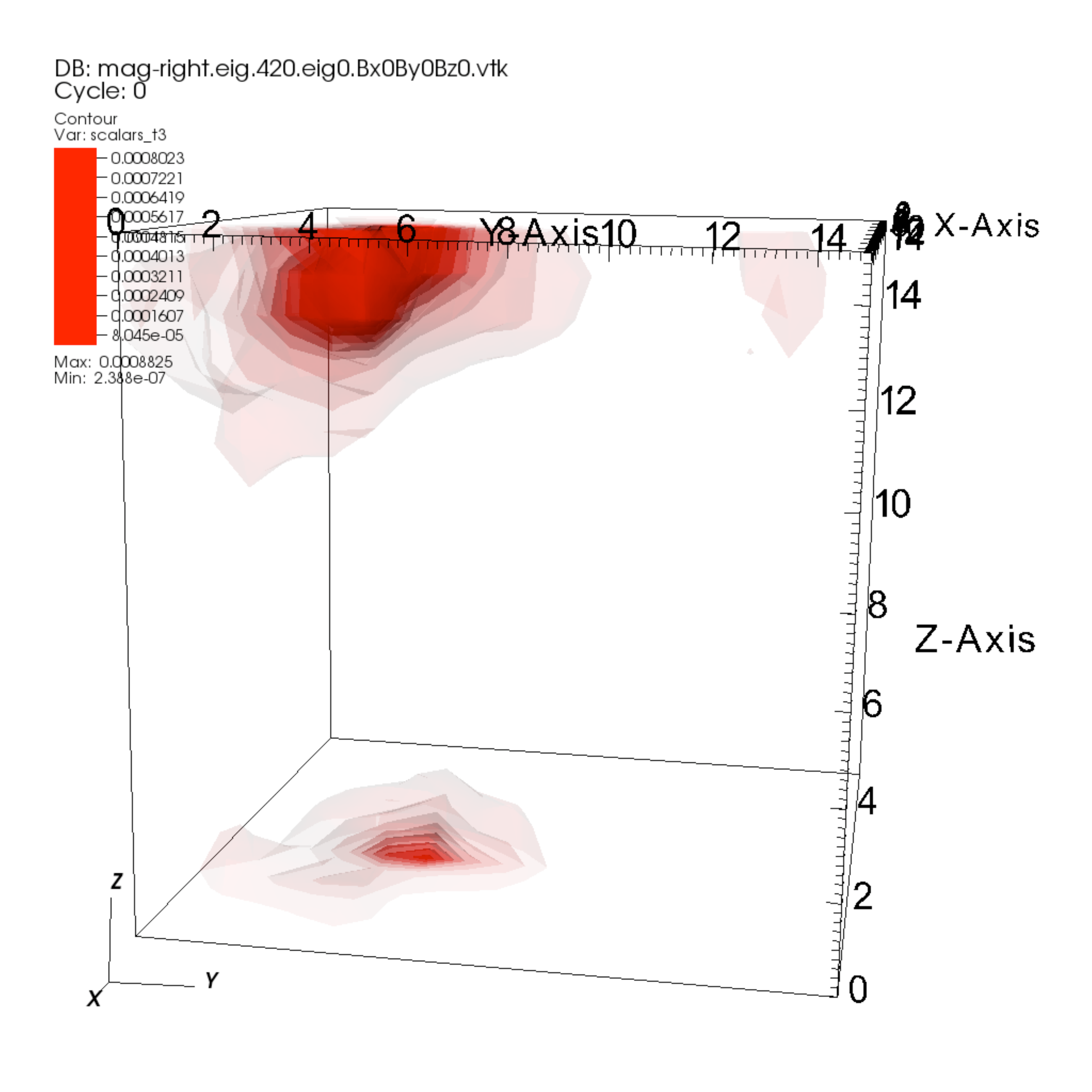}\hskip 0pc
\includegraphics[width=18pc]{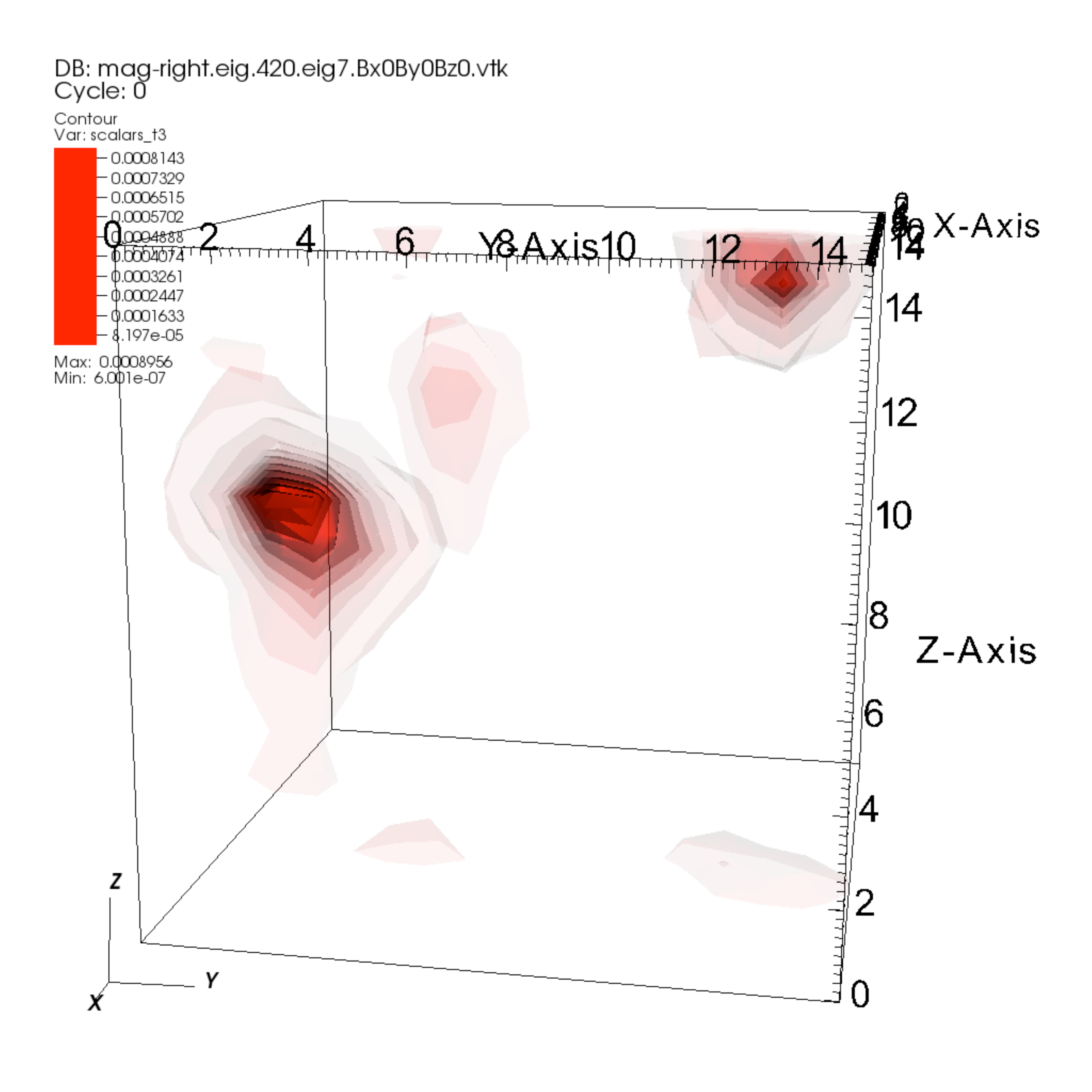}\hskip -0pc
\includegraphics[width=18pc]{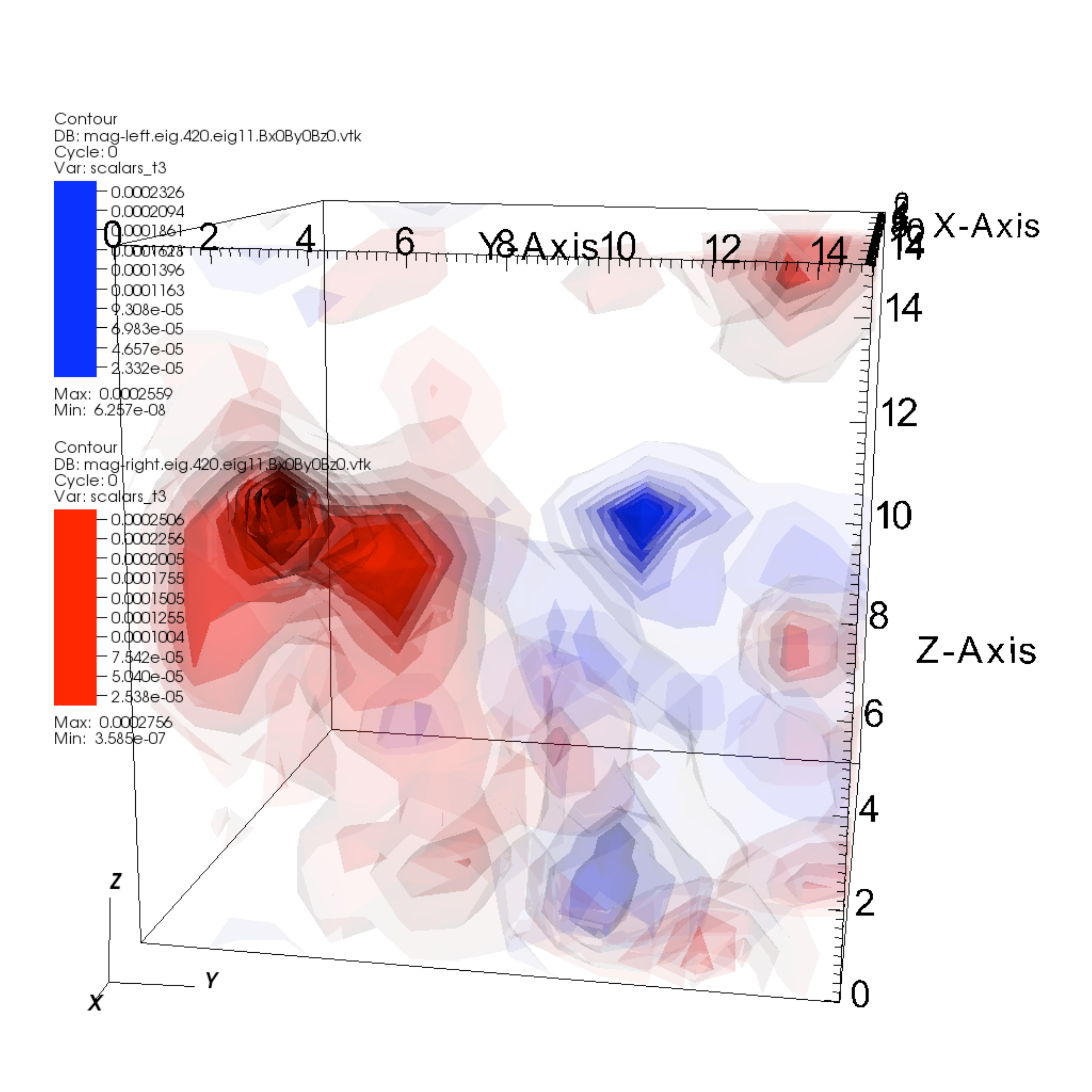}  
  \caption{Top left panel: Topological charge density. Top right: right-handed zero-mode, eigenvector 0. Bottom: right-handed zero-mode, eigenvector 7 (right panel) and eigenvector 11 (left). The eigenmodes are all localized around the ``instantons".  $B_z=0$.}
\label{fig:top charge}
\end{figure}

Having established the chiral magnetic effect in the case of the continuum-like instanton, we turn to the relevant case of 2+1 flavor QCD above $T_c$. We study one configuration from our ensemble, corresponding to monte-carlo time unit 420. Using the 5LI definition of $G\tilde G$ \cite{de Forcrand:1997sq} and 60 steps of APE smearing, the topological charge density for time slice three is shown in the top-left panel of Fig.~\ref{fig:top charge}. Clearly there are several ``lumps" of topological charge and associated localized zero-modes of the Hermitian DWF Dirac operator. Note, all of the zero modes (10) are right-handed on this configuration and were calculated for a valence quark mass of $m_v=0.0001$ in order to expose their continuum like chiral nature. It is interesting to note that the first non-zero mode pair of eigenmodes is localized around an instanton-anti-instanton pair, as shown in the top-left and bottom-right panels of Fig.~\ref{fig:top charge}. Similar results hold for $B_z\neq0$.

\begin{figure}[hbt]
  \includegraphics[width=20pc]{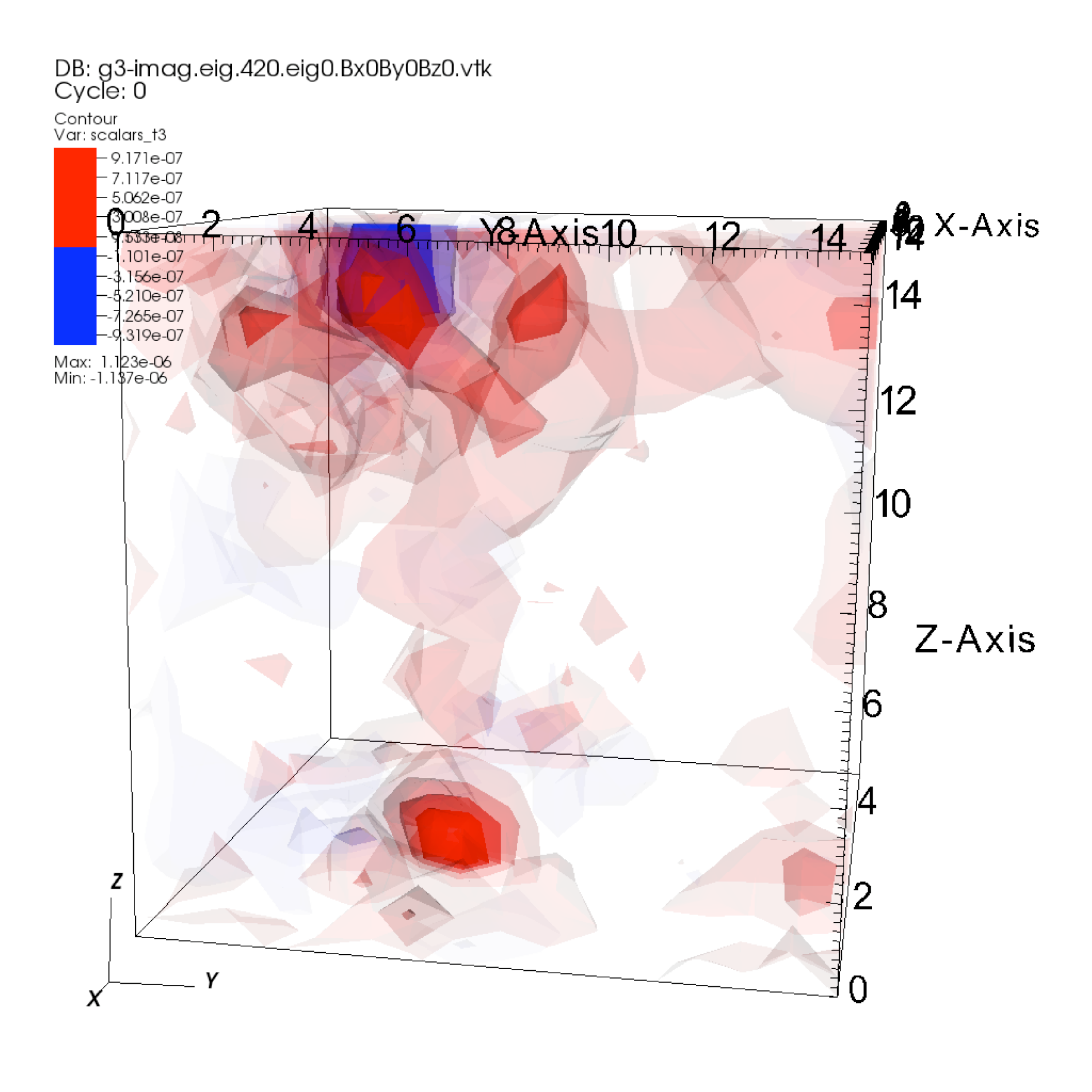}\hskip -2pc
\includegraphics[width=20pc]{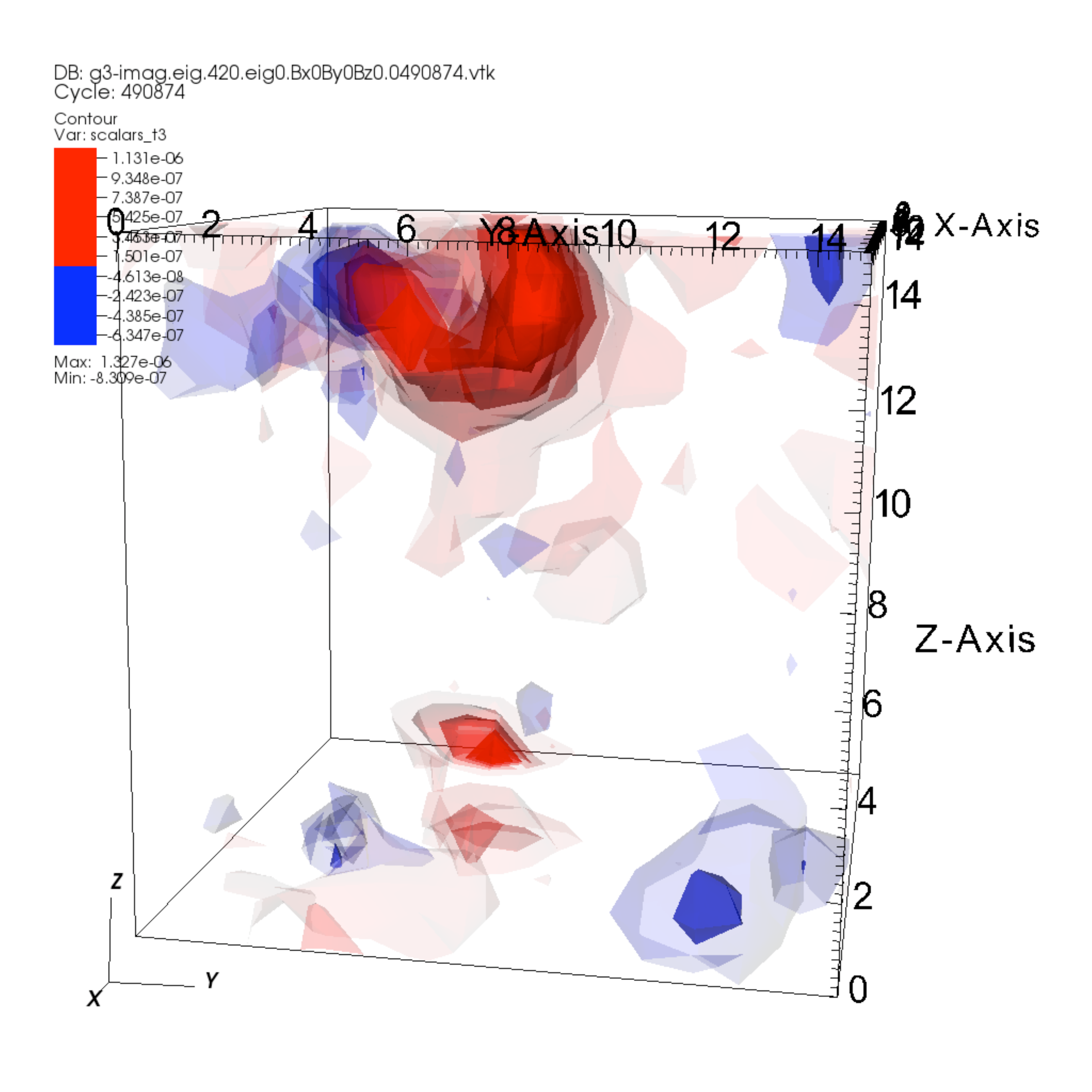}\hskip -2pc\\
\includegraphics[width=20pc]{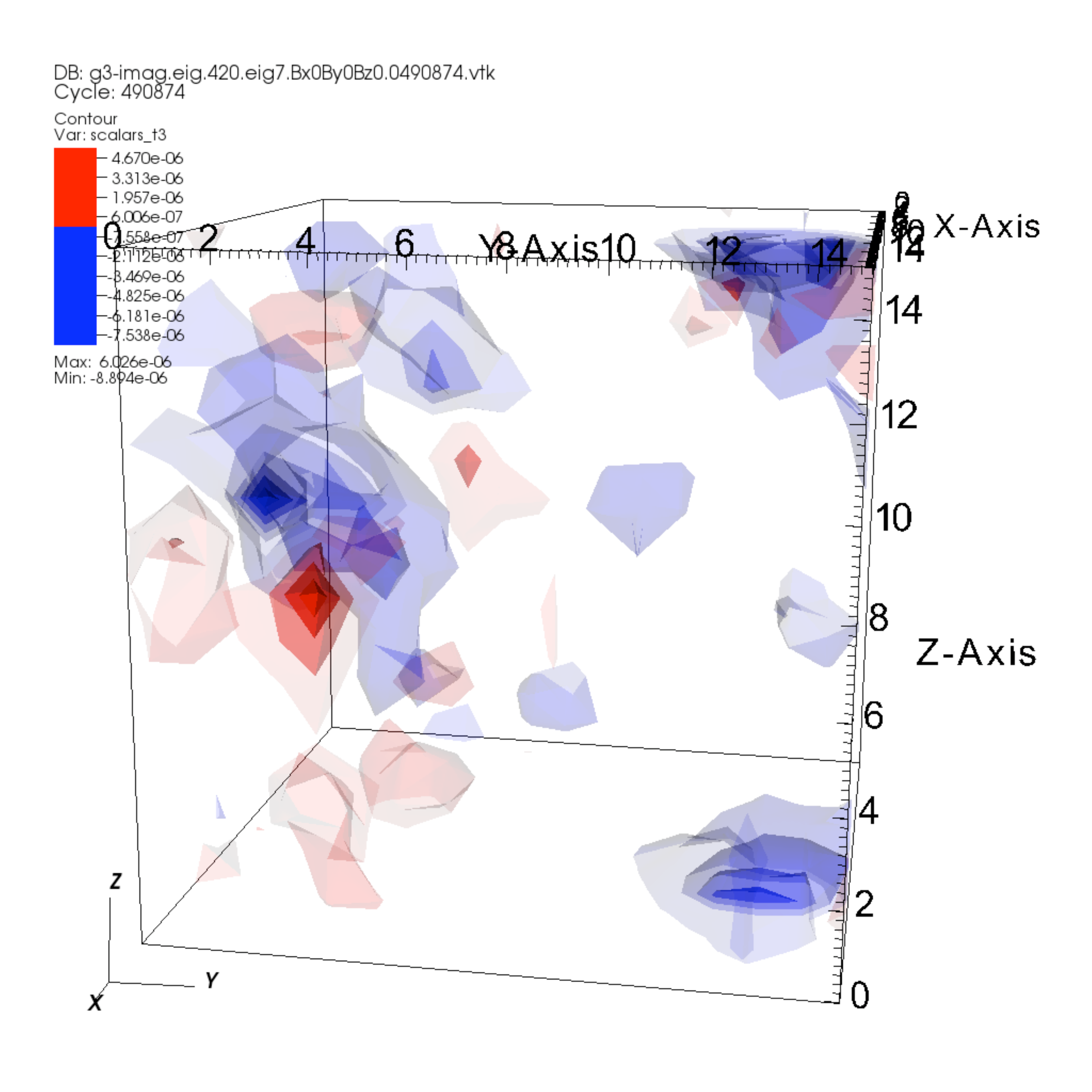}\hskip -2pc
\includegraphics[width=20pc]{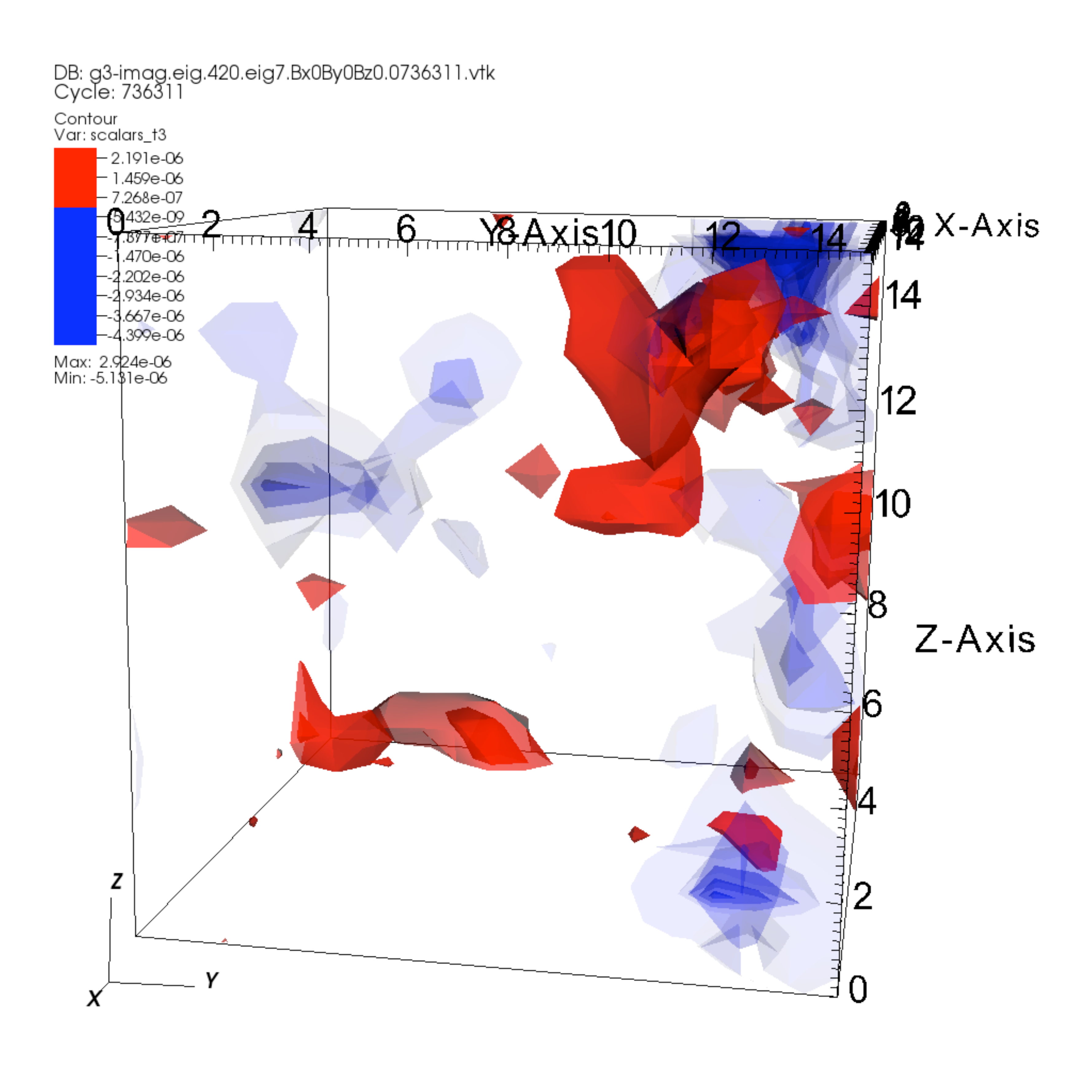}\hskip -2pc
  \caption{Charge density from (near) zero-modes. Top left panel: eigenvector 0, $B_z=0$ (left) and  $B_z=0.0490874$ (right); bottom: eigenvector 7, $B_z=0.0490874$ (left) and $B_z=0.0736311$ (right).}
\label{fig:charge dens}
\end{figure}

Next, we compute the charge density from these zero-modes in the presence of a magnetic field in the $z$-direction, $q B_z=2\pi n /L_x L_y$, with $n=1$, 2, 3, and 4 ($q=1$). In Fig.~\ref{fig:charge dens} the charge density is shown for several values of $B_z$ and two different (would-be) zero-modes. Note, even for $B_z=0$ there is some charge density because the zero-modes are not exactly chiral. It is relative to this ``background" that we seek the charge separation induced by the magnetic field. %; $\bar\psi\gamma_0\psi$ is not zero on a single configuration, only the average vanishes. 
 The peak values appear to be localized around the lumps, possibly suggesting charge is separated above and below the centers of topological charge density shown in Fig.~\ref{fig:top charge} (see the bottom panels in Fig.~\ref{fig:charge dens}). To get at the net charge separated on this configuration, the center of the topological charge distribution in the $z$-direction is determined, $\bar z=\sum z G\tilde G/\sum G \tilde G$. On configuration 420, $\bar z\approx 9.5$. The separated charge is just the total charge in the two halves of the lattice above and ``below" this plane. Unfortunately, the electric charge has been computed using the time component of the local current instead of the point-split conserved one. And the total charge on the lattice in the case where $B_z\neq0$ is not conserved compared to the case where $B_z=0$. This is in marked contrast to the classical instanton case described above. Thus we can not conclude that electric charge has been separated.

\section{Summary} We have reported on a preliminary investigation of the chiral magnetic effect in 2+1 flavor QCD+QED. The effect was clearly established in the case of a classical instanton discretized on a four dimensional lattice. On a more interesting 2+1 flavor QCD+QED configuration above $T_c$ the charge separation around the instantons in an external magnetic field, while suggestive (see Fig.~\ref{fig:charge dens}), was not be clearly observed and needs further investigation. The non-conserved local current was used to compute the charge, and the small symmetry breaking induced by finite $L_s$ was ignored. The results are being updated to address these issues, so we can confirm and understand this exciting discovery made at RHIC. In particular, it is important to determine the dependence on temperature and the strength of the external magnetic field. It will also be interesting to take full advantage of these unique QCD+QED configurations to investigate the correlations of the QED and QCD fields, in particular $(\vec E\cdot \vec B)_{QED}(\vec E\cdot \vec B)_{QCD}$ which we leave for future work.

\vskip0.075in{\noindent\bf Acknowledgements} \vskip.05in 
We thank Dima Kharzeev for helpful discussions and Massimo Di Pierro for help with the 3$d$ graphics. Computations were carried out on the NYBlue BG/L and RBRC QCDOC at BNL, for which we are grateful. T.B. was supported by the US DOE under grant \# DE-FG02-92ER40716.

\end{document}